\documentclass[aps,pre,twocolumn]{revtex4-1}  

\usepackage{bbm,graphicx,bm,bbm,amsmath,hyperref}   

\def\ave#1{\langle #1 \rangle}
\def\cL{{\cal L}}
\def\ii{{\rm i}}
\def\sx{\sigma^{\rm x}}
\def\sy{\sigma^{\rm y}}
\def\sz{\sigma^{\rm z}}
\def\s1{{\mathbbm{1}_2}}

\def\mb{{\bar{\mu}}}
\def\ket#1{{\vline \, #1 \rangle}}
\def\braket#1#2#3{\langle #1 | #2 | #3 \rangle}
\def\1{\mathbbm{1}}

\def\tr#1{{\rm tr}{{#1}}}

\def\etal#1{#1}

\def\tit#1{}

\begin{document}

\title{Large-deviation statistics of a diffusive quantum spin chain and the additivity principle}

\author{Marko \v Znidari\v c}
\affiliation{
Physics Department, Faculty of Mathematics and Physics, University of Ljubljana, SI-1000 Ljubljana, Slovenia}

\date{\today}

\begin{abstract}
Using the large-deviation formalism, we study the statistics of current fluctuations in a diffusive nonequilibrium quantum spin chain. The boundary-driven XX chain with dephasing consists of a coherent bulk hopping and a local dissipative dephasing. We analytically calculate the exact expression for the second current moment in a system of any length and then numerically demonstrate that in the thermodynamic limit higher-order cumulants and the large-deviation function can be calculated using the additivity principle or macroscopic hydrodynamic theory. This shows that the additivity principle can also hold in systems that are not purely stochastic, and can in particular be valid in quantum systems. We also show that in large systems the current fluctuations are the same as in the classical symmetric simple exclusion process.
\end{abstract}



\maketitle

\section{Introduction}

With advancing quantum technologies that are able to coherently manipulate quantum objects the interest in dynamics of quantum systems is increasing. Of particular importance is understanding their nonequilibrium properties. However, there is a fundamental obstacle: while the theoretical formalism for describing equilibrium systems is well known, there is no universal formalism applicable to nonequilibrium situations. In certain simple states, most notable are nonequilibrium steady states (NESSs) to which nonequilibrium systems converge after a long time, a general method though is known. It was originally developed in probability theory and is known as the large-deviation (LD) formalism~\cite{Oono:89,Ellis:85,Varadhan:08,Touchette}.

Provided one is able to calculate the relevant quantities of the LD formalism -- the cumulant generating function and the large-deviation function -- one has access to a full distribution function of an observable, say of a nonequilibrium current. Formal mathematical manipulations involved are actually analogous to equilibrium statistical mechanics which invokes the Legendre transformation to relate different thermodynamic potentials that in turn determine probability distributions. In fact, the mathematical language of statistical mechanics is the LD formalism, although it is usually not presented as such. The very reason that one can speak about intensive thermodynamic quantities, such as temperature, is that in large systems fluctuations are small, which is nothing but a mathematical statement that the LD principle holds. We also note that recently a somewhat related concept called the concentration of measure gained popularity in physics~\cite{Hayden:06}. While the statements there, namely the concentration about the average, are in a sense more general, they give only an upper bound on the probability of fluctuations. The LD theory on the other hand gives a precise quantitative statement, but is valid only asymptotically in the limit of a large sample size (what is a sample size depends on the context; in our case it is the duration of current measurement, in equilibrium statistical mechanics it is usually system size).

The LD approach is well developed for classical equilibrium systems~\cite{Ellis:85,Touchette}. A bit less is known about equilibrium quantum systems. For instance, one tries to rigorously, using $C^*$ algebra, prove the validity of the LD principle, see e.g. Refs.~\cite{KMS}. Treating nonequilibrium systems, for instance NESSs, is more complicated. In general, one expects universal features to appear only in the thermodynamic limit (TDL) and therefore, even though the physics of small systems might be of interest~\cite{small:13}, an especially sought-for are exact LD solutions of large systems. Most exact nonequilibrium calculations of large deviation fluctuations have been done for classical stochastic models, see, e.g., Refs.~\cite{Derrida:98,Derrida:01,Essler:11,Gorissen:12,Lazarescu:13}. Applications of the LD method to quantum systems are more scarce~\cite{Keiji:07,Garrahan:10,Budini:10,Li:11,Lesanovsky:12,Ates:12}. Most deal with small systems, such as one or two qubits~\cite{Garrahan:10,Li:11,Lesanovsky:12}, or the mean field approximation~\cite{Ates:12}. Recently, the LD formalism for the current in a coherent ballistic XX spin chain in the TDL has been provided~\cite{Znidaric:14} (see also~\cite{Medvedyeva:13} for some results) showing an interesting nonanalytic behavior. Nonanalytic behavior of LD functions can be either due to a degenerate NESS~\cite{Garrahan:10,Hurtado:13} or, more interestingly, due to a genuine nonequilibrium phase transition. 

For diffusive systems, provided certain conditions are met, a general theory of nonequilibrium fluctuations can be developed. This is called macroscopic fluctuation theory~\cite{Bertini:01,Bertini:02} and it can for instance predict the probability of observing nonequilibrium density profiles as well as currents~\cite{Bertini:05}. Another useful rule is the additivity principle~\cite{Bodineau:04} that can, similarly as the macroscopic fluctuation theory, predict the LD function based on only the first two cumulants. So far both principles have been verified mostly for classical stochastic models~\cite{Derrida:07}. 

In the present work we provide a solution for current fluctuations in a diffusive quantum model, namely for a driven XX spin chain with dephasing, and we show that the additivity principle holds. The dynamics of the system studied will be governed by the Lindblad equation~\cite{Lindblad},
\begin{eqnarray}
\label{eq:Lin}
\frac{{\rm d}\rho}{{\rm d}t}&=&\ii [\rho,H]+\cL^{\rm dis}\rho,\\
&&\quad \cL^{\rm dis}\rho=\sum_j 2 L_j\rho L_j^\dagger - L_j^\dagger L_j \rho - \rho L_j^\dagger L_j, \nonumber
\end{eqnarray}
where we denote the right-hand-side as a Liouvillian operator $\cL$ acting on a density matrix. We shall calculate current fluctuations in the NESS, which is the state $\rho_\infty$ that is a solution of the stationary Liouville equation $\cL(\rho_\infty)=0$. We also note that most existing solutions for the LD function or cumulants in nonequilibrium quantum systems have been obtained within the Lindblad setting, besides those already mentioned, see also e.g.~\cite{Buca:13,Scholes:13}. Cumulants can also be used to infer unknown Lindblad generators~\cite{Bruderer:13} from a measured dynamics.

\section{Large-deviation formalism}

A large-deviation approach is a systematical mathematical procedure by which we can calculate a distribution function of a sum of random variables in the limit of summing a large number of variables. Provided the central limit theorem is valid, we know that the fluctuations around the average are Gaussian. The large-deviation formalism goes beyond that by providing the whole distribution function, predicting, for instance, also the probability of large fluctuations.

Physicists are actually familiar with the basic steps of the formalism. Namely, the mathematical formalism of equilibrium statistical mechanics, with Legendre transformation relating various thermodynamic potentials, for instance, the entropy and the free energy, is nothing but the large-deviation formalism. The entropy plays the role of a large-deviation function (up to a sign and an additive constant) while the free energy plays the role of a cumulant generating function. Indeed, we know that the entropy determines the probability of a given microstate, similarly to how the large-deviation function determines the probability that a random variable has a certain value. The mathematical details about the large-deviation formalism were worked-out in '60, and its use in statistical physics was ``advertised'' in '80~\cite{Ellis:85,Oono:89}, see also the recent review~\cite{Touchette} for an exposition.

Let us have a look on how the formalism works on a particular observable, which we shall call a current $j(\tau)$, and its closely related integrated quantity, called the number of particles, $N_t=\int_0^t j(\tau)\,{\rm d}\tau$. We shall be interested in a stationary (in general, nonequilibrium) distribution of current, or, equivalently, of the number of transferred particles in time $t$. Time $t$ is our scaling variable (analogous to the number of random variables in a sum, if we were interested in the fluctuation properties of a sum of random variables). The statement of the LD formalism is that if $N_t$ satisfies a so-called LD property then its probability distribution $P(N_t) $ decays exponentially for large $t$, and it can therefore be written as $P(N_t\equiv J^{(t)} t) \sim {\rm e}^{-t\,\Phi(J^{(t)})}$, where $\Phi(J)$ is called a LD function (also a rate function) and $J^{(t)}\equiv N_t/t$ is the average current as determined by counting the number of transferred particles in time $t$. The sign $\sim$ is meant to denote that the relation holds for large $t$ and up to an irrelevant normalization constant that we do not write out. The goal of the LD formalism is to calculate the LD function $\Phi(J)$. The simplest method to get $\Phi$ is via the G\" artner-Ellis theorem. The procedure begins by calculating the moment generating function $\langle {\rm e}^{s N_t}\rangle=\int{{\rm e}^{s N_t}P(N_t){\rm d}N_t}$ (the average is in our nonequilibrium setting over different realizations (measurements) of variable $N_t$) and from it the cumulant generating function $\Lambda(s)$,
\begin{equation}
\Lambda(s)\equiv \lim_{t \to \infty}\frac{1}{t} \ln{\langle {\rm e}^{s N_t}\rangle}.
\label{eq:L}
\end{equation}
The theorem then says that if $\Lambda(s)$ is differentiable for all $s$, then the LD function is given by the Legendre-Fenchel transform of $\Lambda(s)$,
\begin{equation}
\Phi(J)=\max_{s}\{Js-\Lambda(s) \},
\label{eq:LF}
\end{equation}
and the probability distribution for large $t$ is
\begin{equation}
P(J\equiv N_t/t) \sim {\rm e}^{-t\, \Phi(J)}.
\label{eq:LD}
\end{equation}
For strictly convex $\Lambda(s)$ the Legendre-Fenchel transform simplifies to a more familiar Legendre transform $\Phi(J)=s_*J-\Lambda(s_*)$, where $s_*$ is the unique solution of $\Lambda'(s_*)=J$. We see that the crucial step is to calculate the cumulant generating function $\Lambda(s)$. Once we have it, we can calculate $\Phi(J)$ as well as, for instance, the current cumulants. From Eq.~(\ref{eq:L}) we see that the derivatives of $\Lambda(s)$ are cumulants of $N_t$, or, because we have $J^{(t)}=N_t/t$, also appropriately scaled current cumulants,
\begin{equation}
\left. J_r \equiv \frac{{\rm d}^r \Lambda(s)}{{\rm d}s^r}\right|_{s=0} =\lim_{t \to \infty} \frac{1}{t}\langle N_t^r \rangle_{\rm c}=\lim_{t \to \infty} t^{r-1}\ave{[J^{(t)}]^r}_{\rm c} ,
\label{eq:Jr}
\end{equation}
where the subscript ``c'' denotes the cumulant. We shall call the $r$-th derivative of $\Lambda(s)$ simply a (scaled) current cumulant and denote it by $J_r$.

Because one has $N_t=\int_0^t j(\tau)\,{\rm d}\tau$, we immediately see that the $r$-th cumulant of $N_t$ can be expressed in terms of an $r$-point time correlation function of the current $j(\tau)$. Therefore, in order to be able to calculate these cumulants in the NESS, we will have to go beyond just calculating the NESS because $\rho_\infty$ itself only gives a time-independent part of a state. For Markovian processes the cumulant generating function is equal to the largest eigenvalue of a so-called tilted Liouvillian. To see how the tilted Liouvillian is constructed, it is easiest to think in terms of a stochastic unraveling of our Lindblad equation, i.e., in terms of a stochastic wavefunction $\psi(t)$ that exhibits stochastic jumps according to the Lindblad operators involved. The Lindblad dissipator (\ref{eq:Lin}) $\cL^{\rm dis}$ has two types of terms: one is the so-called jump term $2L_j\rho L_j^\dagger$ that in stochastic unraveling induces jumps $\psi(t) \to L_j \psi(t)$, and the other, $- L_j^\dagger L_j \rho - \rho L_j^\dagger L_j$, can be thought of as being part of a non-Hermitean Hamiltonian, see, e.g., Ref.~\cite{Breuer}. In particular, assuming that all Lindblad operators are such that terms involving $L_j^\dagger L_j$ do not change $N_t$ while the jump terms change $N_t$ by either $0$ or $\pm 1$, the tilted Liouvillian $\cL(s)$ is obtained multiplying the jump term by the weight ${\rm e}^{\pm s}$. Specifically, grouping Lindblad operators $L_j$ in three sets: the set $S_0$ of those $L_j$ that do not change $N_t$, the set $S_{+}$ of those $L_j$ that change $N_t$ by $+1$, and the set $S_{-}$ of operators that change $N_t$ by $-1$, we can write the tilted Liouvillian dissipator $\cL^{\rm dis}(s)$ as a sum of three parts, $\cL^{\rm dis}(s)=\cL^{\rm dis}_0(s)+\cL^{\rm dis}_+(s)+\cL^{\rm dis}_{-}$, with
\begin{eqnarray}
\label{eq:til}
\cL^{\rm dis}_{+}(s)&=& \sum_{L_j \in S_+} {\rm e}^{s}\,2L_j\rho L_j^\dagger,\\
\cL^{\rm dis}_{-}(s)&=& \sum_{L_j \in S_{-}} {\rm e}^{-s}\,2L_j\rho L_j^\dagger, \nonumber \\
\cL^{\rm dis}_{0}(s)&=& \sum_{L_j \in S_0} 2L_j\rho L_j^\dagger - \sum_{L_j\in S_0,S_\pm} (L_j^\dagger L_j \rho + \rho L_j^\dagger L_j).\nonumber
\end{eqnarray}
The tilted Liouvillian $\cL(s)$ is obtained by simply using $\cL^{\rm dis}(s)$ instead of $\cL^{\rm dis}=\cL^{\rm dis}(s=0)$ in the Lindblad equation (\ref{eq:Lin}). The cumulant generating function $\Lambda(s)$ is then equal to the largest eigenvalue of $\cL(s)$. To see this, let us decompose $\rho(t)$ into $n$-resolved density matrices $\rho_n(t)$, being the part of $\rho(t)$ that has $N_t=n$. Writing $\rho(t)=\overline{\sum_n |\psi_n(t)\rangle\langle \psi_n(t)|}$ in terms of an average~\cite{Breuer} over stochastic trajectories $|\psi_n(t)\rangle$, where we labeled a trajectory by an index $n$ denoting the value of $N_t$, i.e., being equal to the number of up jumps minus the number of down jumps upto time $t$, we have $\rho(t)=\sum_n \rho_n(t)$~\cite{footrhon}. With $\rho_n(t)$, the probability is simply $P(N_t=n)=\tr{\rho_n(t)}$. Defining an $s$-transformed $\rho_s(t)\equiv \sum_n {\rm e}^{sn}\rho_n(t)$, we also have $\ave{{\rm e}^{sN_t}}=\tr{\rho_s(t)}$. Because $N_t$ is changed only by the jump terms from $S_\pm$, the n-resolved $\rho_n(t)$ satisfies the master equation ${\rm d}\rho_n(t)/{\rm d}t=\cL_0+\cL^{\rm dis}_+(0)\rho_{n-1}(t)+\cL^{\rm dis}_-(0)\rho_{n+1}(t)$. Multiplying the equation by ${\rm e}^{sn}$ and summing over $n$, we finally obtain the equation for $\rho_s(t)$, which is ${\rm d}\rho_s(t)/{\rm d}t=\cL(s)\rho_s(t)$, with the dissipative part of $\cL(s)$ being given by Eq.~(\ref{eq:til}). For large $t$, the norm of $\rho_s(t)={\rm e}^{\cL(s) t}\rho_s(0)$ will grow with a rate given by the eigenvalue of $\cL(s)$ that has the largest real part, and therefore the cumulant generating function $\Lambda(s)$ is equal to the eigenvalue of $\cL(s)$ with the largest real part. Observe that for $s=0$ we have an ordinary non-tilted Lindblad equation and therefore one always has $\Lambda(s=0)=0$. For more details on the LD formalism, see Ref.~\cite{Touchette}.

In mesoscopic physics and quantum optics one is often interested in current cumulants as they can contain information about a system's properties. In '90 a so-called full counting approach was suggested~\cite{Levitov} that is used to calculate cumulants. While the LD and the full counting approach are very similar -- formally, in full counting statistics the weight used is ${\rm e}^{\ii s}$ instead of ${\rm e}^{s}$ -- the analytic properties of the calculated eigenvalue are quite different. For instance, an added bonus in the LD approach is that one gets, besides just cumulants, which by themselves are not very informative, also a full distribution function together with a thermodynamic-like formalism relating $\Lambda(s)$ and $\Phi(J)$. This makes it possible to study the nonanalytic properties of the rate function $\Phi(J)$, for an example see Ref.~\cite{Znidaric:14}, which would not be possible by studying only cumulants. Vaguely speaking, the greater power of the LD formalism lies in sampling stochastic trajectories with an exponentially larger/smaller probability ${\rm e}^{\pm s}$, instead of using just a phase, and thereby inferring the probability of orbits carrying more or less current than on average.

\section{The Model}

The Hamiltonian of our model is given by the XX spin chain,
\begin{equation}
H=\sum_{j=1}^{L-1} \sx_j \sx_{j+1}+\sy_j \sy_{j+1},
\end{equation}
where $\sx,\sy$ are standard Pauli matrices and the chain has $L$ sites. The dissipative part is described by a set of Lindblad operators $L_j$ (\ref{eq:Lin}). For a boundary driven XX chain with dephasing studied here we are going to use $4$ driving Lindblad operators $L_j$ at chain ends, i.e., at the first and the last site, and $L$ dephasing Lindblad operators, one for each lattice site. Lindblad operators representing driving are given by the following expressions,
\begin{eqnarray}
\label{eq:Ls}
\!\! L_1=\sqrt{\Gamma_{\rm L}(1+\mu+\mb)}\,\sigma^+_1&,& L_2=\sqrt{\Gamma_{\rm L}(1-\mu-\mb)}\, \sigma^-_1\\
\!\! L_3=\sqrt{\Gamma_{\rm R}(1-\mu+\mb)}\,\sigma^+_L&,& L_4=\sqrt{\Gamma_{\rm R}(1+\mu-\mb)}\, \sigma^-_L,\nonumber
\end{eqnarray}
with $\sigma_k^{\pm}=(\sx_k\pm \ii\, \sy_k)/2$. Dephasing Lindblad operator at site $j$ is on the other hand 
\begin{equation}
L^{\rm deph}_j=\sqrt{\frac{\gamma}{2}} \sz_j,
\label{eq:Ldeph}
\end{equation}
where $\gamma$ is the dephasing strength. $L$ dephasing Lindblad operators (\ref{eq:Ldeph}) together with $4$ boundary ones (\ref{eq:Ls}) constitute a set of operators $L_j$ in the Lindblad equation (\ref{eq:Lin}). We are interested in current statistics in the NESS state $\rho_\infty$. While the NESS itself can be calculated exactly~\cite{Znidaric:10,Znidaric:11} (because of a decoupled hierarchy of correlations, see also Refs.~\cite{eisler:11,temme:12}) here we are going to study current fluctuations, which requires us to solve a more complicated tilted-Liouvillian equation.

Let us briefly explain the role of all parameters. $\Gamma_{\rm L,R}$ are coupling strengths between the chain and the bath at the left/right chain end. Most of the time we shall have the same coupling at both ends, $\Gamma_{\rm L}=\Gamma_{\rm R}\equiv\Gamma$. One should be aware that in order for the measured quantities (e.g., the current or the magnetization profile) to reflect the system's bulk properties, the coupling $\Gamma$ in a one-dimensional boundary driven system should be of order $1$. In particular, one should not use small $\Gamma \to 0$ as this limit is singular; for perturbative results for the XXZ spin chain see~\cite{Prosen:11} (for instance, for small $\Gamma$ the current scales as $\sim \Gamma$ while the magnetization scales as $\sim \Gamma^2$, resulting in a diverging transport coefficient, regardless of the true nature of bulk conductivity). $\gamma$ is the dephasing strength whose nonzero value causes the scaling of the current $j \sim 1/L$ with the system's size and linear magnetization profile in the bulk~\cite{Znidaric:10}, both being characteristic features of diffusive systems. Four driving Lindblad operators (\ref{eq:Ls}) try to induce the average magnetization $\mu+\mb$ at the left chain end and $\mb-\mu$ at the right end, therefore, $\mb$ is the average magnetization in a stationary state and $\mu$ is a nonequilibrium driving parameter. If $\mu=0$, we have an equilibrium driving; if $\mu\neq 0$, we will get a nonequilibrium steady state with a linear magnetization profile. The allowed values of $\mu$ and $\mb$ are such that all the square-roots in four $L$'s (\ref{eq:Ls}) are real. Without sacrificing generality, we can limit to $\mu \in [0,1]$ and $\mb \in [-1,1]$.

To study large deviation statistics of the magnetization current in a stationary state, we need a tilted generator $\cL(s)$ (\ref{eq:Ls}). We are going to measure the size of the transferred magnetization at the right chain end. To achieve that, we add a factor $e^s$ with $L_4$ and a factor $e^{-s}$ with $L_3$, i.e., we have sets $S_+=\{L_4 \}$, $S_{-}=\{L_3\}$, and $S_0=\{L_1,L_2,L_j^{\rm deph}\}$. Note that, because we use a factor ${\rm e}^s$ instead of ${\rm e}^{2s}$ in our definition of $\cL(s)$ (after one application of, say, $\sigma^-_L$, the expectation value of $\sz_L$ changes by $2$, not by $1$), we are in fact studying the particle current $j_k$,
\begin{equation}
j_k\equiv \sx_k \sy_{k+1}-\sy_k \sx_{k+1},
\end{equation}
and not the true magnetization current, being equal to $j_k^{\rm mag}=2j_k$. Cumulant $\tilde{J}^{\rm mag}_r$ of the magnetization current can be simply obtained as $\tilde{J}^{\rm mag}_r=2^r J_r$. To sum-up, in order to evaluate LD statistics, we need the largest eigenvalue $\Lambda(s)$ of the tilted Liouvillian (\ref{eq:til},\ref{eq:Lin}). 

\section{Second current cumulant}

We shall first analytically calculate the exact expressions for the 2nd current cumulant in equilibrium and out of equilibrium. It turns out that this 2nd moment scales as $\sim 1/L$ with system size, similarly as the 1st cumulant, i.e., the average current~\cite{Znidaric:10,Znidaric:11}. The model is diffusive, and it has been argued that for such models one can use a so-called additivity property to get higher-order cumulants just from knowledge of the first two. The additivity has so-far been demonstrated for classical models such as various exclusion models. In the next section, we shall use the 2nd moment calculated here to confirm the validity of the additivity principle also for a diffusive quantum spin chain.  

\subsection{Equilibrium current fluctuations ($\mu=0$)}

We are going to use perturbation theory in the tilting parameter $s$ in order to analytically calculate the 2nd current cumulant, i.e., current fluctuations.

In equilibrium when $\mu=0$ the exact NESS solution at $s=0$ is very simple and equal to $\rho_0=\prod_j (\1+\mb \sigma_j^z)$~\cite{Znidaric:11}. For small $s$ we can make an expansion of the largest eigenvalue and the corresponding eigenvector of $\cL(s)$ as $\Lambda(s)=\Lambda_0+s\Lambda_1+\frac{s^2}{2}\Lambda_2+\cdots$, and $\rho(s)=\rho_0+s\rho_1+\frac{s^2}{2}\rho_2+\cdots$. The Liouvillian $\cL(s)$ can be expanded as well, $\cL(s)=\cL_0+s\cL_1^{\rm R}+\frac{s^2}{2}\cL_2^{\rm R}+\cdots$, where, because $\cL(s)$ depends on $s$ only via the Liouvillian $\cL^{\rm R}$ of the right bath (involving Lindblad operators $L_{3,4}$), we have simple explicit forms of $\cL_1^{\rm R}$ and $\cL_2^{\rm R}$,
\begin{eqnarray}
\cL^{\rm R}_1&=&
2\Gamma_{\rm R}
\begin{pmatrix}
0 & 0 & 0 & 0\\
0 & 0 & 0 & 0\\
0 & 0 & -(\mu-\mb) & -1\\
0 & 0 & 1 & \mu-\mb
\end{pmatrix}, \nonumber \\
\cL^{\rm R}_2&=&
2\Gamma_{\rm R}
\begin{pmatrix}
0 & 0 & 0 & 0\\
0 & 0 & 0 & 0\\
0 & 0 & -1 & -(\mu-\mb)\\
0 & 0 & \mu-\mb & 1
\end{pmatrix}.
\end{eqnarray}
The matrices are written in the basis $\{\sx,\sy,\sz,\1\}$ and we wrote-out only the nontrivial part acting on the last $L-$th spin. Noting that for the largest eigenvalue one always has $\Lambda_0=0$, we get two lowest order perturbative equations,
\begin{eqnarray}
\label{eq:pert}
\cL_0\, \ket{\rho_1}+\cL^{\rm R}_1\,\ket{\rho_0}&=&\Lambda_1\,\ket{\rho_0}\\
\cL^{\rm R}_1\,\ket{\rho_1}+\frac{1}{2}\cL^{\rm R}_2\,\ket{\rho_0}+\frac{1}{2}\cL_0\,\ket{\rho_2}&=&\Lambda_1\,\ket{\rho_1}+\frac{1}{2}\Lambda_2\,\ket{\rho_0}.\nonumber
\end{eqnarray}
Projecting on $\langle \1|$ (i.e., taking a trace) we get the equation for the eigenvalue correction, $\Lambda_1=\braket{\1}{\cL_0}{\rho_1}+\braket{\1}{\cL^{\rm R}_1}{\rho_0}$. The first term is always zero because $\cL_0=\cL(s=0)$ is trace preserving, while the 2nd term is zero for equilibrium $\rho_0$ as one can readily check using the explicit form of $\rho_0$. Therefore, of course, $\Lambda_1=0$, i.e., the average current in equilibrium is zero, $J_1=d\Lambda/ds=\Lambda_1=0$. With this, the eigenvalue equation for $\rho_1$ simplifies in equilibrium to
\begin{equation}
{\cal L}_0\,\ket{\rho_1}=-\cL^{\rm R}_1\,\ket{\rho_0}.
\label{eq:x1}
\end{equation}
Solving it, we can then use $\rho_1$ in the expression for the 2nd order correction to the eigenvalue, obtained by projecting the 2nd equation in Eq.~(\ref{eq:pert}),
\begin{equation}
\Lambda_2=\braket{\1}{\cL^{\rm R}_2}{\rho_0}+2\braket{\1}{\cL^{\rm R}_1}{\rho_1}.
\label{eq:lam2}
\end{equation}
To summarize, in order to obtain the 2nd cumulant $J_2=\Lambda_2$ in equilibrium, we need to first solve Eq.(\ref{eq:x1}) and then evaluate the eigenvalue correction in Eq.(\ref{eq:lam2}). The first term in Eq.(\ref{eq:lam2}) is easy to evaluate using the explicit form of $\rho_0$ and $\cL_2^{\rm R}$, and it is $\braket{\1}{\cL^{\rm R}_2}{\rho_0}=2\Gamma_{\rm R}(1-\mb^2)$. Because $\cL^{\rm R}_1$ acts nontrivially only on the last site, to get the 2nd term in (\ref{eq:lam2}) we only need coefficients of $\1$ and $\sz_L$ in $\rho_1$, in fact, only of $\sz_L$ because $\rho_1$ is orthogonal to $\ket{\1}$ (all $\rho_{j>0}$ are traceless). Denoting this coefficient by $z_L^{(1)}$, Eq.~(\ref{eq:lam2}) gives $\Lambda_2=2\Gamma_{\rm R}(1-\mb^2)+4\Gamma_{\rm R} z_L^{(1)}$. To get $z_L^{(1)}$, we have to solve for $\rho_1$. First, we can calculate $\cL^{\rm R}_1\,\rho_0=-2\Gamma_{\rm R}(1-\mb^2)\sz_L \prod_j^{L-1}(\1+\mb \sz_j)$. The ansatz for $\rho_1$ can be expressed as a linear combination of $\sz_k$ and their products on any number of sites, of the current $j_k$, and of products of one $j_k$ and $\sz_k$ (again, any number of $\sz$'s). The equations are in fact similar to those solved in Ref.~\cite{Znidaric:10} (but slightly more complicated) for the NESS. Solving for magnetization profile terms and the current in $\rho_1$, we get
\begin{equation}
z_L^{(1)}=-(1-\mb^2)\left( \frac{1}{2}-\frac{1}{4\Gamma}\frac{1}{\Gamma+\frac{1}{\Gamma}+\gamma (L-1)}\right),
\end{equation}
from which the 2nd order eigenvalue correction can be calculated, finally obtaining,
\begin{equation}
J_2^{(0)}=\Lambda_2=\frac{2(1-\mb^2)}{\Gamma_R+\Gamma_L+\frac{1}{\Gamma_L}+\frac{1}{\Gamma_R}+2\gamma(L-1)}.
\label{eq:J2eq}
\end{equation}
For later convenience, we denoted the equilibrium fluctuations by $J_2^{(0)}\equiv J_2(\mu=0)$. For large system size $L$, we have the asymptotic form
\begin{equation}
J_2^{(0)} \asymp \frac{1-\mb^2}{\gamma L}.
\label{eq:J2eqAsymp}
\end{equation}

One could continue with the perturbation series to higher orders, however, terms get complicated and we were not able to obtain an exact closed expression. Due to the symmetry of the $\mu=0$ driving, one immediately knows that all odd cumulants are exactly zero, $J_{1,3,5,\ldots}=0$. In the next section, we are going to argue that higher even moments are in general all nonzero for general $\mb$. As we shall see, they can actually be calculated from $J_2^{(0)}$. There are two special points though, for which the behavior is different. One is at $\mb=0$, when, on average, there are as many spins pointing up as down (half-filling in particle language), and for which we shall demonstrate that all higher cumulants are zero in the TDL. The other special case is that of maximal driving, $\mb=1$ (one necessarily also has $\mu=0$), for which all cumulants (including the 1st and the 2nd) are exactly zero because both reservoirs simultaneously try to only inject particles at both ends (an equivalent situation arises also for $\mb=-1$). 

We also observe that in equilibrium, the largest eigenvalue $\Lambda(s)$ (and only the largest one) is invariant under the mapping $\Gamma \to 1/\Gamma$, implying also the invariance of the rate function $\Phi(J)$ under that mapping. For $\gamma=0$, all eigenvalues of $\cL(s)$ are invariant under such mapping, see Ref.~\cite{Znidaric:14}.

\subsection{Non-equilibrium current fluctuations ($\mu\neq 0$)}

Here we shall again use perturbation theory, similarly as in the equilibrium case. For simplicity, we shall use $\Gamma_{\rm L}=\Gamma_{\rm R}=\Gamma$ throughout this part. From the solution for the NESS~\cite{Znidaric:10,Znidaric:11}, we already know the first cumulant,
\begin{equation}
J_1=\Lambda_1=\frac{2\mu}{\Gamma+\frac{1}{\Gamma}+\gamma(L-1)},
\label{eq:J1}
\end{equation} 
as well as the lowest order term $\rho_0$ of the eigenvector $\rho(s)$. From $\rho_0$ one can calculate $\braket{\1}{\cL^{\rm R}_2}{\rho_0}=2\Gamma(z_L^{(0)}(\mu-\mb)+1)$, which is needed in $\Lambda_2$, where we denoted $z_L^{(0)}=\mb-\mu+\frac{\mu \Gamma}{\Gamma (1+\Gamma^2)+\gamma(L-1)}$. To get other terms in the 2nd order correction $\Lambda_2$, we need to solve Eq.~(\ref{eq:pert}) for $\rho_1$. Because $\rho_0$ out-of-equilibrium is more complicated than the one for $\mu=0$ we were not able to get a closed symbolic solution for an arbitrary $L$. For small $L<5$, however, the linear system of equations given by Eq.~(\ref{eq:pert}) can be solved symbolically. While the expression for $L=2$ is special~\cite{foot1}, the expressions for $L=3$ and $L=4$ can already give us some clue as to what the result should look like for general $L$. Unfortunately, the expressions are still sufficiently complicated so that we were not able to write the generic result. We therefore used the numerical solution of perturbative equations for a few larger $L$'s to guide ourselves towards the correct exact form of $J_2$ holding for any $L>2$. The exact result is rather lengthy, and we give it in Appendix~\ref{app:0}. $J_2$ is for all allowed values of $\mu$ and $\mb$ never zero, except in the equilibrium case of $\mb=\pm 1$ and $\mu=0$, when all current cumulants are trivially zero. For the out-of-equilibrium $J_2$ (\ref{eq:J2}), as opposed to the equilibrium $J_2^{(0)}$ (\ref{eq:J2eq}), the symmetry $\Gamma \to 1/\Gamma$ is no longer exact. In fact, in the TDL one has $J_2(\Gamma)-J_2(1/\Gamma) \asymp \frac{2(\Gamma^4-1)}{\gamma^3\Gamma^2 L^4}$. 

While having the exact expression for $J_2$ is nice, we are mainly interested in the behavior in the TDL. For large $L$ only the leading $L$ dependence of Eq.~(\ref{eq:J2}) can be retained, giving the asymptotic dependence
\begin{equation}
J_2 \asymp \frac{3(1-\mb^2)-\mu^2}{3\gamma L}.
\label{eq:J2asymp}
\end{equation}
The Fano factor in the TDL is given by $J_2/J_1\asymp \frac{3(1-\mb^2)-\mu^2}{3\mu}$, and it is independent of $\gamma$ and $\Gamma$.

\section{Full current distribution}

\subsection{The additivity principle}

Calculating higher cumulants gets increasingly more complicated, and a generic, system-independent method would be highly desired. For classical stochastic diffusive models, it has been conjectured that the so-called additivity principle holds~\cite{Bodineau:04}, and this can in turn be used to calculate the cumulant generating function and all higher cumulants only from the first two. The additivity principle for the LD function $\Phi(J)$ is the following assertion: denoting by $\Phi_L(J,n_{\rm L},n_{\rm R})$ the large deviation function for a system of length $L$ with driving that induces density $n\equiv (\sigma_z+1)/2$ equal to $n_{\rm L}$ at the left and $n_{\rm R}$ at the right chain end, in the TDL limit the LD function should satisfy
\begin{equation}
\Phi_{L+L'}(J,n_{\rm L},n_{\rm R})={\rm min}_n[\Phi_{L}(J,n_{\rm L},n)+\Phi_{L'}(J,n,n_{\rm R})].
\label{eq:additivity}
\end{equation}
Because probability is given as $\sim {\rm e}^{-t\Phi}$, the additivity principle (\ref{eq:additivity}) means that the probability in a longer chain is an appropriate maximization of a product of probabilities of two shorter chains. Remember that the LD function plays a role analogous to the entropy in equilibrium physics and so it is in a way natural to be additive if correlations are not too strong, which is what one expects for a diffusive system. Provided additivity holds, it is a very powerful principle. Splitting the chain into many pieces, so that density differences eventually become small, and assuming that the cumulant function scales as $\sim 1/L$, one can use linear response and obtain a general expression for $\Phi(J)$ and $\Lambda(s)$. It depends only on two parameters, namely the diffusion constant $D(n)$ and equilibrium fluctuations $\sigma(n)$ defined in the linear response regime by,
\begin{equation}
J_1=D(n)\frac{\Delta n}{L},\quad J_2^{(0)}=\sigma(n)\frac{1}{L}.
\label{eq:D}
\end{equation}
Both $D(n)$ and $\sigma(n)$ can be in general density-dependent. In particular, the 3rd and 4th current cumulants out-of-equilibrium are~\cite{Bodineau:04}
\begin{equation}
L J_3=\frac{3(I_3 I_1-I_2^2)}{I_1^3},\quad L J_4=\frac{3(5I_4I_1^2-14I_1I_2I_3+9I_2^3)}{I_1^5},
\label{eq:II}
\end{equation}
where 
\begin{equation}
I_k\equiv \int_{n_{\rm R}}^{n_{\rm L}}D(n)\sigma(n)^{k-1}\,{\rm d}n.
\end{equation}

\begin{figure}[t!]
\centerline{\includegraphics[width=8cm]{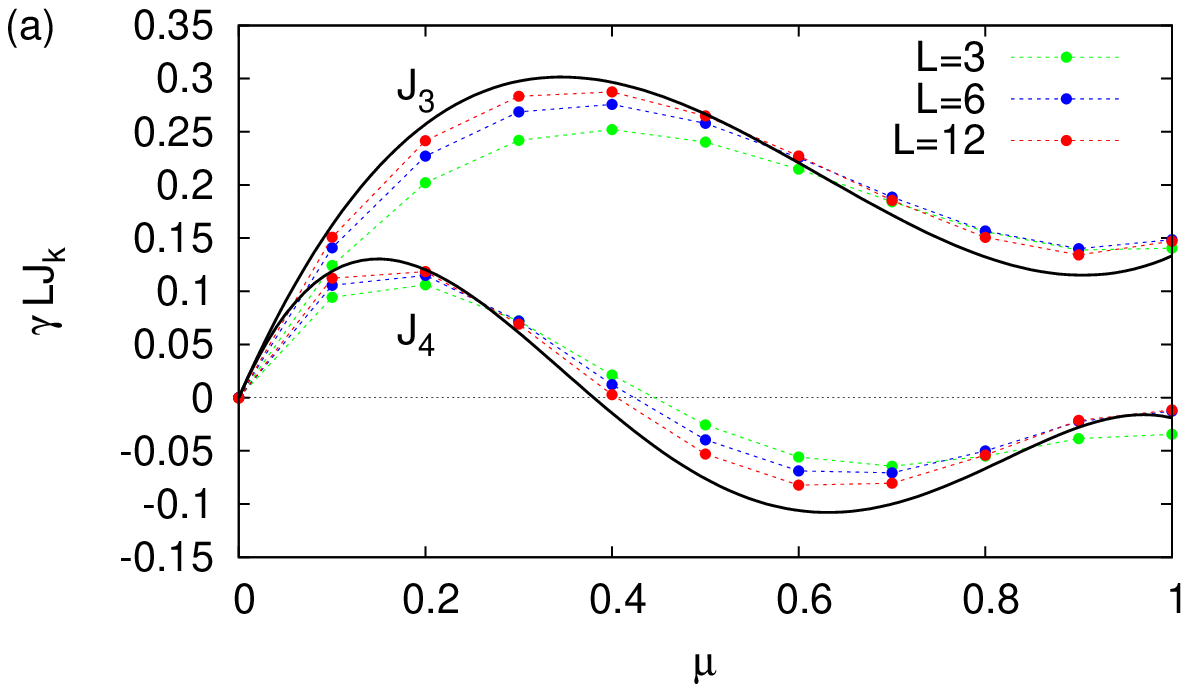}}
\centerline{\includegraphics[width=8cm]{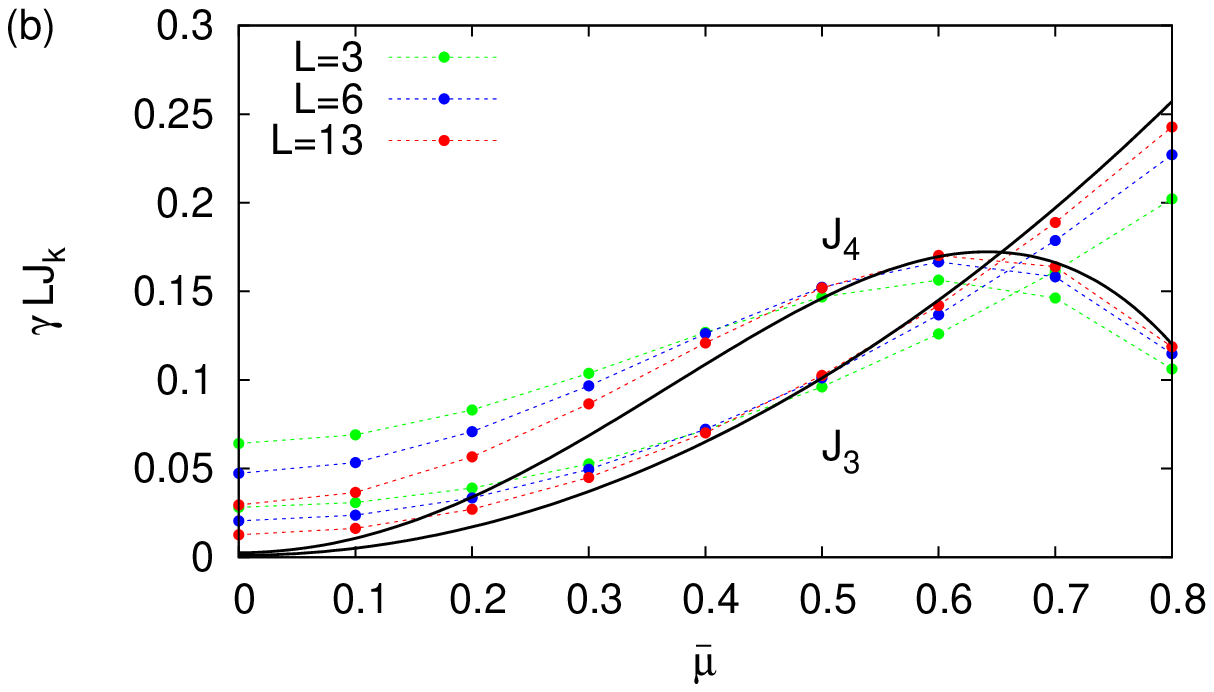}}
\caption{(Color online) Convergence of numerically computed $J_3$ and $J_4$ in the XX dephasing chain towards the additivity prediction (\ref{eq:J34}) shown with full curves. In (a) we use $\mu+\mb=1$, in (b) $\mu=0.2$, and in both cases $\gamma=\Gamma=1$.}
\label{fig:cmax}
\end{figure}
The additivity principle has so far been successfully verified in classical stochastic processes~\cite{Derrida:07}, such as the symmetric simple exclusion process~\cite{Bodineau:04} or the KPZ model~\cite{Hurtado:09}. Its validity is less clear in coherent systems and in quantum systems, see though Ref.~\cite{Saito:11} for a study of a disordered harmonic lattice. The quantum XX chain with dephasing studied here differs from classical stochastic models in two aspects: (i) the hopping term that causes transport of magnetization is coherent and not stochastic, (ii) diffusive transport is due to bulk dephasing described by a dissipative Lindblad term. Therefore, the transport mechanism is more complicated than in exclusion models, however, the dynamics in bulk is still not completely coherent. Namely, ultimately one would like to understand fluctuations in a completely coherent nonequilibrium model. We are going to demonstrate that the additivity principle holds in the quantum XX chain with dephasing by verifying that in the TDL the cumulants agree with the ones predicted (\ref{eq:II}) by the additivity principle. We are also going to check that the full LD function agrees with the one predicted by the additivity principle.

\begin{figure}[t!]
\centerline{\includegraphics[width=8cm]{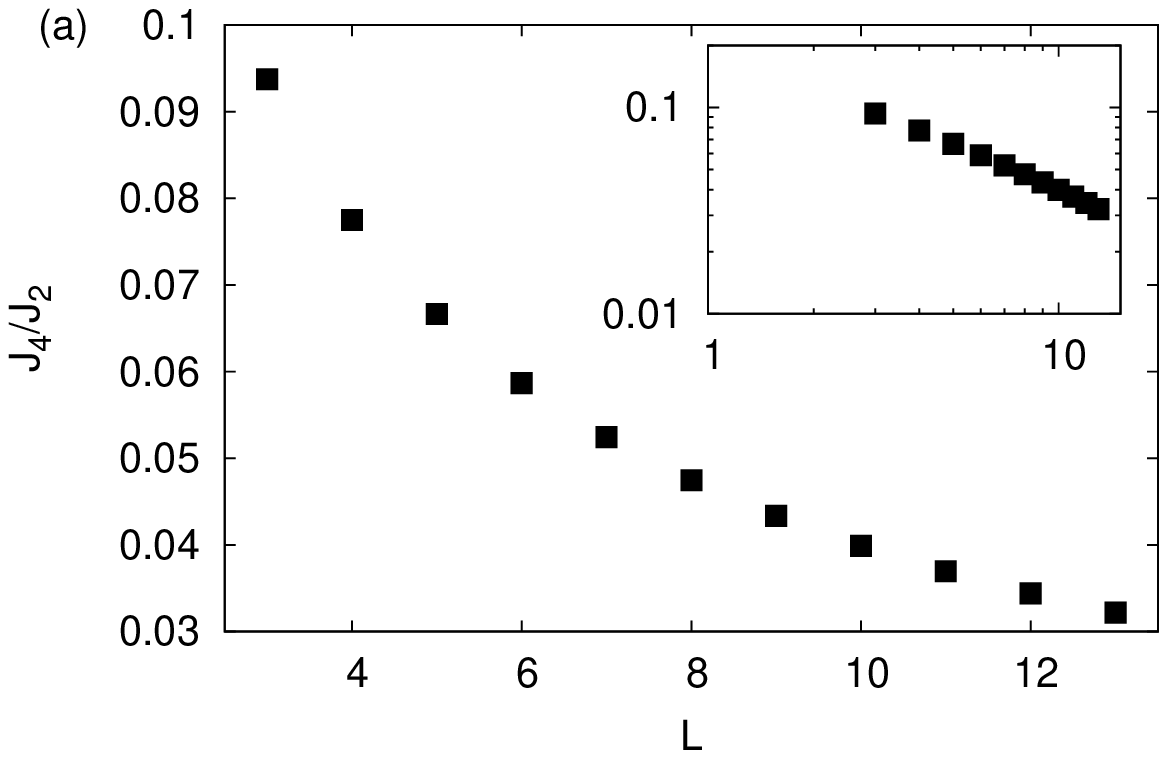}}
\centerline{\includegraphics[width=8cm]{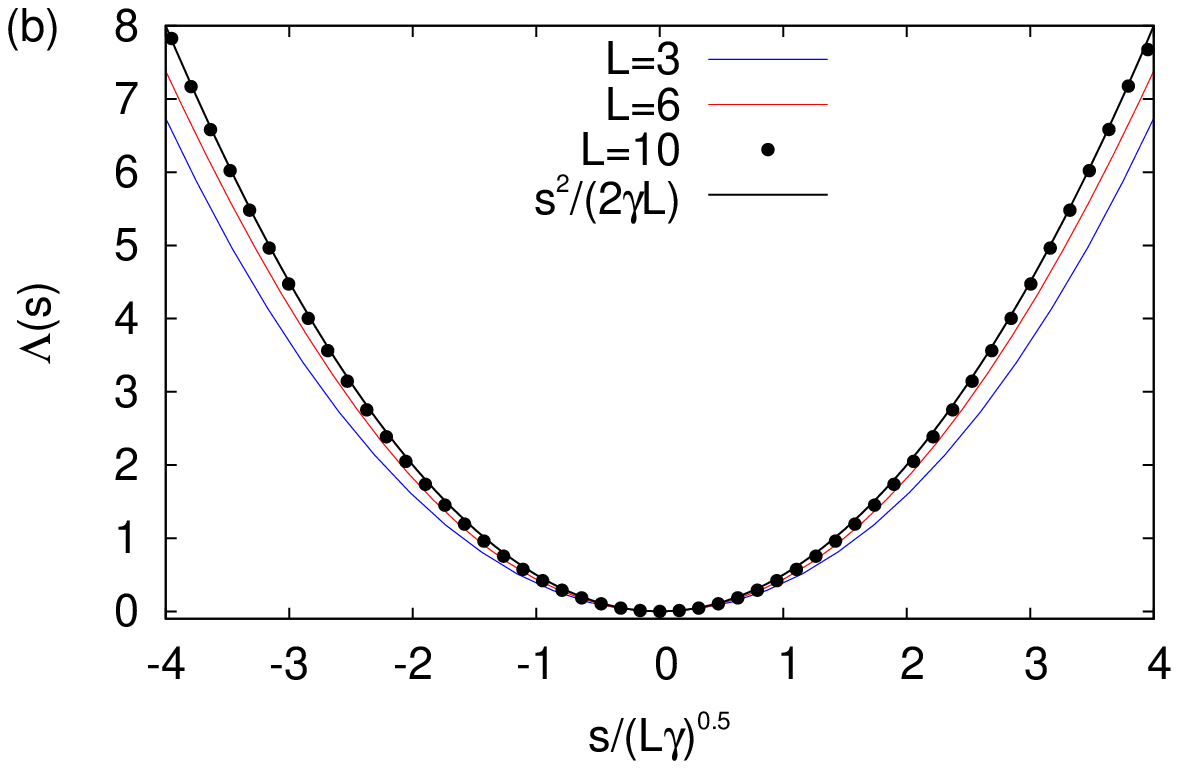}}
\caption{(Color online) For equilibrium driving $\mu=0$ and zero average magnetization $\mb=0$ fluctuations are Gaussian (for any other driving parameters they are not). (a) The ratio $J_4/J_2$ decays to $0$ in the thermodynamic limit. The inset show the same data on a log-log scale. (b) Cumulant generating function $\Lambda(s)$ converges to theoretical Gaussian (\ref{eq:J2eqAsymp}) (full black curve). We use $\Gamma=\gamma=1$ and $\mb=\mu=0$.}
\label{fig:J4eq}
\end{figure}
Using our asymptotic results for $J_2^{(0)}$ (\ref{eq:J2eqAsymp}) and $J_1$ (\ref{eq:J1}), and definitions (\ref{eq:D}), we easily get the two needed parameters,
\begin{equation}
D(n)=\frac{2}{\gamma},\quad \sigma(n)=\frac{4n(1-n)}{\gamma},
\label{eq:Ds}
\end{equation}
where we use density $n=\langle 1+\sz\rangle/2=(1+\mb)/2$ instead of magnetization. Calculating cumulants using (\ref{eq:II}), we in turn get
\begin{eqnarray}
\label{eq:J34}
J_3&=&\frac{2\mu(\mu^2+15\mb^2)}{15\gamma L},\\
J_4&=&\frac{105\mb^2(1-\mb^2)+\mu^2(7-462\mb^2)-9\mu^4}{105\gamma L}.\nonumber
\end{eqnarray}
We have verified these two expressions against numerically calculated cumulants using exact diagonalization (Appendix~\ref{app:A}) for different values of $\mu$ and $\mb$, always finding agreement within finite-size effects. Two representative examples are shown in Fig.~\ref{fig:cmax}. One can see that with increasing system size $L$ the results indeed converge to Eq.~(\ref{eq:J34}). Based on that, we can say that the additivity principle holds for a nonequilibrium XX chain with dephasing. It is worth noting that for some parameter values the convergence can be quite fast, e.g., for $\mu=0.2, \mb=0.5$, while for others it is slower. Also, for generic parameter values cumulants are nonzero, except in the special case of $\mu=\mb=0$. We would also like to point out that in diffusive systems, an example is our XX chain with dephasing, all current cumulants scale as $\sim 1/L$ with the system size.

\subsection{Cumulant generating function}
\begin{figure}[t!]
\centerline{\includegraphics[width=8cm]{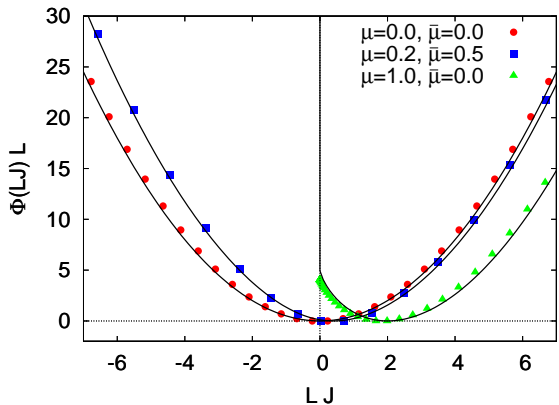}}
\caption{(Color online) Large deviation function for different driving $\mu$, and $\mb$. Points are obtained from numerically calculated $\Lambda(s)$ for $L=10$, while full curves are the asymptotic $L \to \infty$ theory obtained from Eq.~(\ref{eq:Lambda}). We use $\Gamma=\gamma=1$.}
\label{fig:Phi}
\end{figure}

The expressions for $D(n)$ and $\sigma(n)$ (\ref{eq:Ds}) for the XX chain with dephasing are in fact, up to an irrelevant overall time-scale prefactor $2/\gamma$, the same as for classical symmetric simple exclusion process~\cite{Spohn:83,Derrida:04} (SSEP). Because the additivity principle holds also for our quantum model, the whole cumulant generating function $\Lambda(s)$ in the TDL should be equal to the asymptotic cumulant generating function of the SSEP. Using a known form of $\Lambda(s)$ for the SSEP~\cite{Derrida:04,Bodineau:04}, we can write down the cumulant generating function of the Lindblad driven XX dephasing chain,
\begin{eqnarray}
\label{eq:Lambda}
&&\Lambda(s) \asymp \frac{2}{\gamma L}\left[{\rm arcsinh} \left( \frac{1}{2}\sqrt{x}\right)  \right]^2, \\
x&=&(1-{\rm e}^{-s})\left[ \mb^2-(1-\mu)^2\right]+(1-{\rm e}^s)\left[\mb^2-(1+\mu)^2\right].\nonumber
\end{eqnarray}
This expression, holding in the TDL, generates the same cumulants as the additivity principle (\ref{eq:II}). Note that the XX spin chain with dephasing is not a direct quantum analog of the SSEP; see Refs.~\cite{eisler:11,temme:12} for examples of ``quantum'' exclusion models. In particular, the evolution includes a coherent part given by the Hamiltonian as well as an incoherent dephasing part. 

The cumulant generating function (\ref{eq:Lambda}) is generally non-Gaussian, causing also the fluctuations to be non-Gaussian. The only special point for which fluctuations are Gaussian is in equilibrium $\mu=0$ at zero average magnetization $\mb=0$ (i.e., at half-filling). This can also be seen from the additivity principle~\cite{Bodineau:04}: fluctuations are generally Gaussian only for equilibrium driving, $n_{\rm L}=n_{\rm R}=n_*$, where $n_*$ is such that $\sigma(n)$ is maximal at $n_*$ (from (\ref{eq:Ds}) we see that in our system $n_*=1/2$). We have numerically checked that fluctuations are indeed Gaussian for $\mu=\mb=0$ in the TDL by calculating $J_4$ and $\Lambda(s)$, see Fig.~\ref{fig:J4eq}.

From the cumulant generating function, either the numerically calculated one or the asymptotic exact result (\ref{eq:Lambda}), one can use the Legendre transform (\ref{eq:LF}) to obtain the LD function $\Phi(J)$. In Fig.~\ref{fig:Phi} we compare numerical results for $L=10$ with the prediction of the asymptotic theory obtained by Legendre transforming Eq.~(\ref{eq:Lambda}). One can see that $\Phi(J)$ has a zero at the most probable value of the nonequilibrium current, that is at $J_1$ (\ref{eq:J1}). Around the zero the shape of $\Phi(J)$ is parabolic, signifying the Gaussian nature of small fluctuations, while for larger $|J-J_1|$ there can be deviations from a parabolic shape. In particular, for $\mu+\mb=1$ (or in the equivalent situation of $\mu+\mb=-1$) the current can only be positive (only injection of magnetization at one chain end) which is reflected in an infinite value of $\Phi(J)$ for $J<0$ and therefore the strong non-Gaussian nature of large fluctuations.

For one-dimensional diffusive systems, a general approach has been developed called macroscopic fluctuation theory~\cite{Bertini:01,Bertini:02}. Using just the two input parameters $D(n)$ and $\sigma(n)$, macroscopic fluctuation theory can be used to calculate the large deviation functional of a nonequilibrium density. Remember that obtaining large deviation results for the density is in general considerably more difficult than for the current. We also note that, provided certain conditions are met~\cite{Bertini:05}, the additivity principle can be derived from the macroscopic fluctuation theory. It would be interesting to check the validity of macroscopic fluctuation theory for the density in the model studied here.

\begin{figure}[t!]
\centerline{\includegraphics[width=8cm]{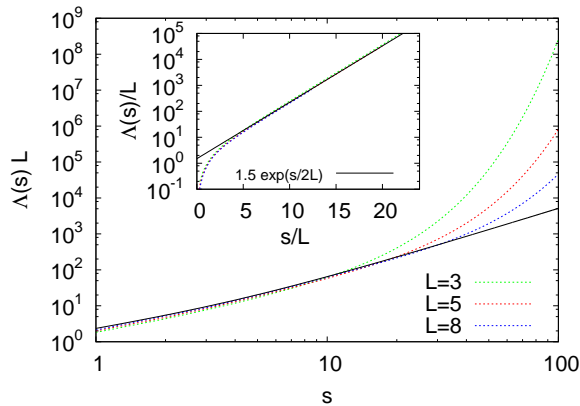}}
\caption{(Color online) Finite-size effect at maximal driving $\mu=1$ and $\mb=0$. Dashed colored curves are numerical results, the full black curve is the exact result in the TDL, Eq.~(\ref{eq:Lambda}), growing as $\sim s^2$ for large $s$. Inset: finite-$L$ numerical data (overlapping dashed colored curves) grow with $s$ exponentially (full black line). We use $\gamma=\Gamma=1$.} 
\label{fig:Lambda1L}
\end{figure}

\subsection{Finite-size effects}

We have already discussed what are the finite-size effects for cumulants. For $\Lambda(s)$ another effect is important. Namely, it turns out that for finite $L$, the behavior of $\Lambda(s)$ for large $s$ is eventually dominated by finite-size effects. Let us demonstrate this by a simple example, taking parameters $\mu=1$ and $\mb=0$. In this special case, the cumulant function (\ref{eq:Lambda}) simplifies to $\Lambda(s)L\gamma = 2[{\rm arcsin}(\sqrt{{\rm e}^s-1})]^2$. It is worth noting that this expression is the same as for a disordered quantum conductor at low temperatures, obtained by averaging Landauer single-channel results over the universal distribution of transmission coefficients~\cite{Lee:95} (and it is of course the same as for the SSEP~\cite{Derrida:07}). For large positive $s$, the cumulant generating function therefore behaves in the TDL as $\Lambda(s)L\gamma \asymp 2(\ln{2}+\frac{s}{2})^2$, meaning that large positive current fluctuations are Gaussian distributed, even though smaller fluctuations are not. For any finite $L$ though, $\Lambda(s)$ will asymptotically grow exponentially with $s$ and not as $\sim s^2$. This can be seen in Fig.~\ref{fig:Lambda1L}. We can see that for sufficiently large $s$, the largest eigenvalue grows as $\Lambda(s) \asymp \frac{3}{2}L\exp{(\frac{s}{2L})}$ (for $\gamma=\Gamma=1$). Such exponential asymptotic growth happens for $s> s_{\rm c}$, where one can estimate that $s_{\rm c} \propto L$. In the TDL, this exponential growth is pushed to infinity and the behavior predicted by Eq.~(\ref{eq:Lambda}) is recovered. Such finite-size effects for large $s$ are generic, see also the results in Ref.~\cite{Znidaric:14}.

\section{Conclusion}

We studied nonequilibrium current fluctuations in a driven quantum spin chain that displays diffusive transport. We analytically calculated the second current moment and then numerically showed that higher order moments, as well as the large deviation function, can be calculated using the additivity principle. They are equal as in the classical symmetric simple exclusion process. The presented results are a first step towards understanding nonequilibrium fluctuations in large coherent quantum systems. Namely, the dynamics of the studied model is a combination of coherent evolution and dissipation, and one nevertheless observes the validity of the additivity principle and of macroscopic fluctuation theory that have been confirmed so far mostly in classical (stochastic) models. It would be interesting to see if those principles hold also in purely coherent quantum models. The presented findings could also serve as a benchmark on how fluctuations in diffusive quantum systems out-of-equilibrium behave -- for instance, that all cumulants are inversely proportional to system length, as opposed to ballistic systems, where they are all independent of system length~\cite{Znidaric:14}.

\appendix

\section{Exact expression for $J_2$}
\label{app:0}
\begin{figure}[th!]
\centerline{\includegraphics[width=8cm]{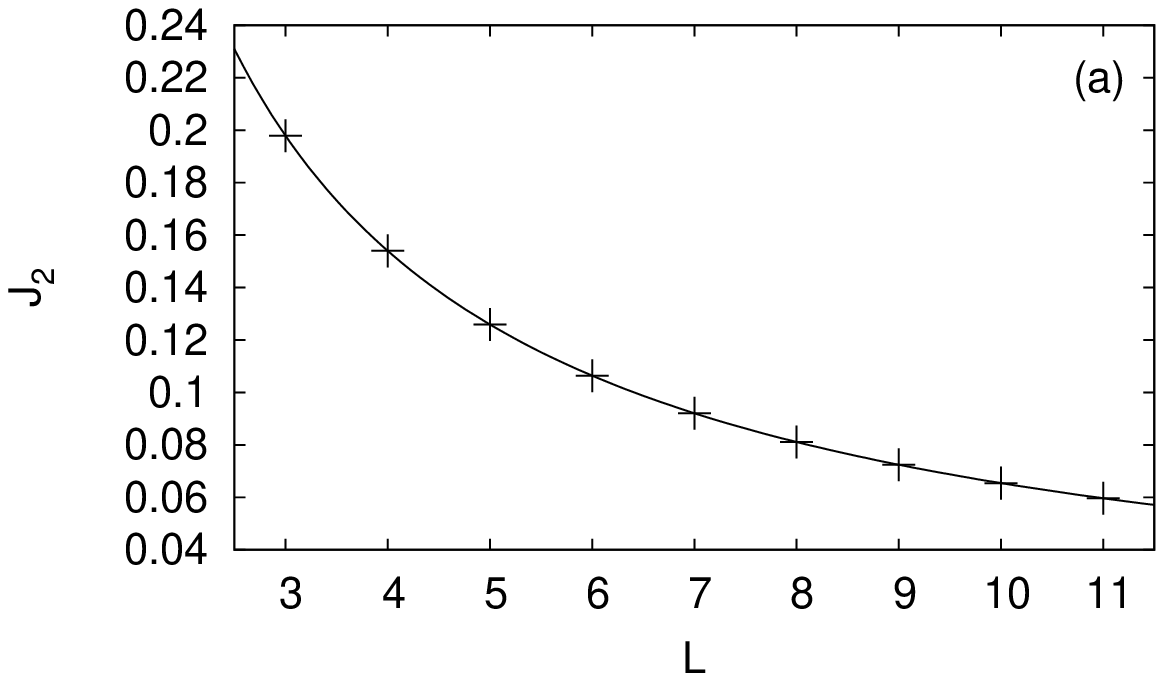}}
\centerline{\includegraphics[width=8cm]{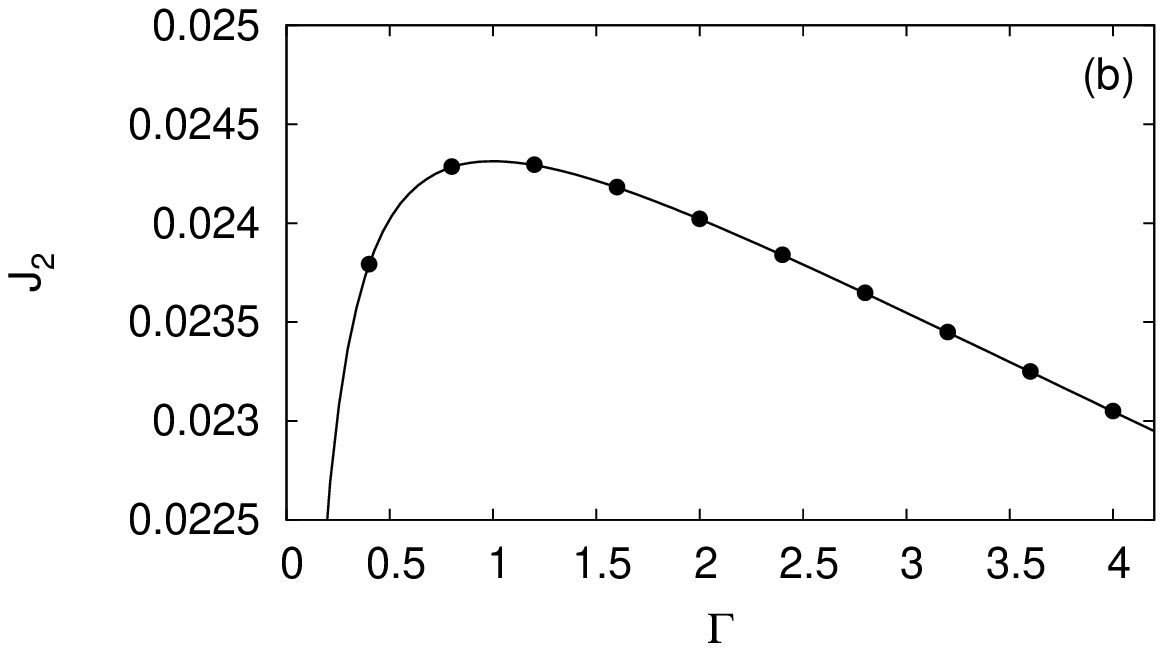}}
\caption{Current fluctuations $J_2$ for the XX dephasing model. Full curve is in both figures the theory given by Eq.~(\ref{eq:J2}). a) points are numerical $J_2$ obtained by exact diagonalization; data are for $\mu=1$, $\gamma=\Gamma=1$, and $\mb=0$, for which the theory (full curve) simplifies to $J_2=(3+L+6L^2+2L^3)/(3L(1+L)^3)$. b) Dependence on $\Gamma$ for $L=40$, $\gamma=1$, $\mu=0.1$, and $\mb=0$. Points represent tDMRG data. Observe that the amplitude of variation with $\Gamma$ is a sub-leading $\sim 1/L^2$ correction.}
\label{fig:J2}
\end{figure}
The exact expression for the 2nd current cumulant in a nonequilibrium XX chain with dephasing is given by,
\begin{equation}
J_2=J_2^{(0)}-\mu^2\Gamma^2\frac{(L-1)\gamma+a(\Gamma+\Gamma^3)+b \Gamma^2+c\Gamma^4}{D[1+\Gamma^2+(L-1)\gamma\Gamma]^3},
\label{eq:J2}
\end{equation}
where $a=3+(L-1)(L-3)\gamma^2$, $b=\gamma[ 3L-7+\frac{1}{3}(L-1)(L-2)(L-3)\gamma^2]$, $c=(L-3)\gamma$, and $D=1+(L-2)\gamma\Gamma+\Gamma^2$.
The $J_2^{(0)}$ in the above expression is the exact equilibrium 2nd cumulant given by Eq.~(\ref{eq:J2eq}). To really make sure that Eq.~(\ref{eq:J2}) is indeed exact we have checked it against numerically calculated cumulants using either exact diagonalization (for $L<12$) or tDMRG simulations for longer chains, see Appendix~\ref{app:A} and \ref{app:B} for details on both numerical methods. For instance, in Fig.~\ref{fig:J2} we compare Eq.~(\ref{eq:J2}) with numerical results of exact diagonalization and of tDMRG simulation, seeing a perfect agreement.

\section{Writing $\cL(s)$ as a ladder}
\label{app:A}

The tilted generator $\cL(s)$ acts on a Hilbert space of operators. For spin-$1/2$ chains a local operator basis is of dimension $4$ and is therefore isomorphic to a Hilbert space of two spin-$1/2$ particles. If $\cL(s)$ acts in a nearest-neighbor fashion on a lattice of length $L$ we can organize these two particles spanning a local operator basis into a rung and write the whole $\cL(s)$ as a spin ladder of length $L$. 

\begin{figure}[ht!]
\centerline{\includegraphics[width=8cm]{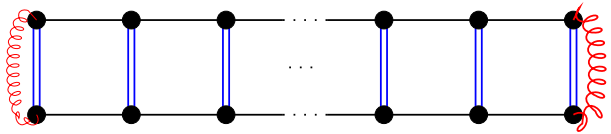}}
\caption{(Color online) Schematic diagram of the tilted Liouvillian $\tilde{\cL}(s)$ for a boundary driven XX chain with dephasing, expressed as a non-Hermitean ladder. Double lines along rungs (blue) are $\sz \tau^{\rm z}$ coupling due to dephasing, two springs at the two boundaries (red) are due to boundary driving that involves tilting by ${\rm e}^s$ at the right end, see Eq.(\ref{eq:ladder}) for details.}
\label{fig:ladder}
\end{figure}

Mapping the local operator basis $\{\sx,\sy,\sz,\1\}$ to a rung basis $\{\ket{00},\ket{10},\ket{01},\ket{11}\}$ by the prescription
\begin{eqnarray}
\ket{\sx}\to \ket{00}+\ket{11}&,&\quad \ket{\sy}\to \ii(\ket{00}-\ket{11}), \nonumber \\
\ket{\sz}\to \ket{01}-\ket{10}&,&\quad \ket{\1}\to \ket{01}+\ket{10},
\end{eqnarray}
which can also be compactly written as the mapping $|\phi \rangle \langle \psi | \to \ket{\phi} \otimes \sx \ket{\psi}$~\cite{Prosen:12}, the resulting form of $\cL(s)$ is a rather simple non-Hermitean spin-$1/2$ ladder. Equivalently, doing local rotation by $U_L\equiv U^{\otimes L}$, $\tilde{\cL}\equiv U_L \cL(s) U_L^\dagger$, where
\begin{equation}
U=\frac{1}{\sqrt{2}}\begin{pmatrix}
1 & \ii & 0 & 0\\
0 & 0 & -1 & 1\\
0 & 0 & 1 & 1\\
1 & -\ii & 0 & 0
\end{pmatrix},
\end{equation}
written again in basis $\{\sx,\sy,\sz,\1\}$, we get (Fig.~\ref{fig:ladder})
\begin{widetext}
\begin{eqnarray}
\tilde{\cL}=&&\ii\{H(\sigma)-H(\tau)\}-\gamma\sum_{j=1}^L \sz_j\tau^{\rm z}_j+2\Gamma_{\rm R} \left\{\frac{\mb-\mu}{2}(\sz_L-\tau^{\rm z}_L)+{\rm e}^{-s}(1+\mb-\mu)\sigma_L^+ \tau_L^- + {\rm e}^{s}(1+\mu-\mb)\sigma_L^- \tau_L^+  \right\}+\nonumber \\
&&+2\Gamma_{\rm L} \left\{\frac{\mu+\mb}{2}(\sz_1-\tau^{\rm z}_1)+(1+\mu+\mb)\sigma_1^+ \tau_1^- +(1-\mu-\mb)\sigma_1^- \tau_1^+  \right\}-(\gamma L+2\Gamma_{\rm R}+2\Gamma_{\rm L})\,\1,
\label{eq:ladder}
\end{eqnarray}
\end{widetext}
where $\sigma_j$ and $\tau_j$ are Pauli matrices on the 1st and 2nd leg, respectively, and $H(\sigma)$ is the XX chain on the 1st ladder leg while $H(\tau)$ is the XX chain on the 2nd leg. The cumulant generating function $\Lambda(s)$ is equal to the largest eigenvalue of $\tilde{\cL}$. We observe that the operator $\tilde{\cL}$ (\ref{eq:ladder}) commutes with total magnetization $\sum_{j=1}^L \sigma_j^{\rm z}+\tau_j^{\rm z}$ and therefore the eigenvalue problem has a block structure (there are other discrete symmetries that we shall not exploit). The largest eigenvalue $\Lambda(s)$ that we seek is always from the sector with zero total magnetization. Using this symmetry reduces for large $L$ the Hilbert space size on which we have to diagonalize $\tilde{\cL}$ by a factor of $\sim \sqrt{L\pi}$ from total size $4^L$. 

To get $\Lambda(s)$ one can use exact diagonalization for small $L$ ($L \lesssim 8$), while for slightly larger $L$ ($L \lesssim 12\sim 13$) the Arnoldi method is better. To get cumulants from numerically calculated $\Lambda(s)$ we used a finite difference approximation of derivatives. Using the difference of $\Delta s=0.01$ usually gives cumulants with enough precision for our purposes.

\section{tDMRG calculation of $\Lambda(s)$}
\label{app:B}
The time-dependent density matrix renormalization group (tDMRG) method (sometimes also called time-evolved-block-decimation) is a method by which one can calculate ground states of one-dimensional quantum systems as well as do time evolution. It can also be extended for a simulation of the time evolution of Lindblad master equations, and in particular for the calculation of NESSs, see Ref.~\cite{JSTAT:09} and a detailed description in Ref.~\cite{Znidaric:10c}. Part of the algorithm for Lindblad equations is periodic Schmidt re-orthogonalization of the state $\rho(t)$. While for unitary evolution re-orthogonalizations are not necessary, non-unitary terms present in Lindblad equations destroy Schmidt decomposition and also the optimality of the method. In the orthogonalization procedure, one also checks for Schmidt orthogonality, being a condition on matrices $M_j^{\nu_j}$ describing a matrix product operator ansatz for $\rho(t)$ (the notation we use here is the same as in the Appendix of Ref.~\cite{Znidaric:10c}). In particular, one should have (see Eq.~(A.7) in Ref.~\cite{Znidaric:10c})
\begin{equation}
r_j^{(k)}\equiv |\sum_{\nu_j,p} [M_j^{\nu_j}]_{k,p} [M_j^{\nu_j}]^*_{k,p}|=1,
\label{eq:r}
\end{equation}
for each site $j$ and each matrix index $k$. While the tilted $\cL(s)$ is not trace-preserving anymore, the very same tDMRG method that is used to calculate ${\rm e}^{\cL(0)t}\rho(0)$ can nevertheless be used to also calculate $\Lambda(s)$. The idea is very simple: for long times the norm of ${\rm e}^{\cL(s)t}\rho(0)$, and with it also the norm of matrices $M_j^{\nu_j}$, will grow due to a positive largest eigenvalue $\Lambda(s)$. Factors $r_j^{(k)}$ will therefore not be $1$ anymore but slightly larger. To get $\Lambda(s)$ one therefore has to remember the values of $r_j^{(k)}$ before every re-normalization of $M_j^{\nu_j}$. Specifically, provided that $r_j^{(k)}$ increased from $1$ to $r_j^{(k)}\approx 1+2\epsilon_j^{(k)}$ in some short time $dt$ (after long time, when $\rho(t)$ converges), the largest eigenvalue is $\Lambda(s)=\frac{1}{dt}\sum_{j=1}^L \epsilon_j^{(k)}$.

We have used this method to calculate $\Lambda(s)$ as well as cumulants by using finite differences to approximate derivatives. There are though some limitations. To calculate higher cumulants with satisfactory precision, one needs $\Lambda(ds)$ with high precision. High precision of $\Lambda(s)$ quickly translates into a high dimension of matrices and therefore slow simulation. In practice, the method can be easily used to calculate 2nd cumulants (see the results in Fig.~\ref{fig:J2}), and with much greater effort also the 3rd and 4th in some cases. Higher cumulants are probably out of reach. To calculate $\Lambda(s)$ for large $s$ one has to face another problem. Because $\Lambda(s)$ increases, a correspondingly smaller time-step must be used in the simulation. In addition, sometimes the method experiences convergence issues that we think might be due to eigenvalues of $\cL(s)$ with large complex parts, causing unwanted oscillations. Perhaps a better approach would be to use tDMRG directly on $\tilde{\cL}$ and search for the ground state of a non-Hermitean $\tilde{\cL}$. Such an approach has been used in Ref.~\cite{Gorissen:11} for a classical exclusion process.


\begin{thebibliography}{}
\bibitem{footrhon} Alternatively, to obtain $\rho_n(t)$ one can add an inert ``environment ancilla'' $|N_t\rangle $ that counts $N_t$. Extending Lindblad operators from sets $S_\pm$ by $L_j \to L_j \otimes \sigma^\pm_{\rm anc.}$, having no Hamiltonian evolution on the ancilla subspace, and starting evolution on the extended Hilbert space with $\tilde{\rho}(0)=\rho(0)\otimes |0\rangle\langle 0|$, we can see that one always has the diagonal form $\tilde{\rho}(t)=\sum_n \rho_n(t)\otimes |n\rangle\langle n|$ as well as $\rho(t)={\rm e}^{\cL t}\rho(0)={\rm tr}_{\rm anc.}\tilde{\rho}(t)$.
\bibitem{foot1} The case of $L=2$ is different because there are no bulk spins, and one has $J_2=J^{(0)}_2-\mu^2\Gamma^2\frac{\gamma+(3-\gamma^2)\Gamma-2\gamma\Gamma^2-\Gamma^3}{(1+\Gamma^2+\gamma\Gamma)^3}$.
\end{thebibliography}
\end{document}